\documentclass[a4paper,11pt]{article}
\pdfoutput=1 

\usepackage{jcappub} 

\usepackage[T1]{fontenc} 

\title{\boldmath Stability of a 
tachyon braneworld}

\author[a]{Gabriel Germ\'an,}
\author[b,c]{Alfredo Herrera-Aguilar,}
\author[a,d]{\\ Andr\'e Martorano Kuerten,}
\author[a]{Dagoberto Malag\'on--Morej\'on,}
\author[e]{\\ and Rold\~ao da Rocha}

\affiliation[a]{Instituto de Ciencias F\'isicas, Universidad Nacional Aut\'onoma de M\'exico,\\
Apartado Postal 48-3, 62251, Cuernavaca, Morelos, M\'{e}xico}
\affiliation[b]{Instituto de F\'{\i}sica, Benem\'erita Universidad Aut\'onoma de Puebla,\\ 
Apartado Postal J-48, 72570, Puebla, Puebla, M\'exico}
\affiliation[c]{Instituto de F\'{\i}sica y Matem\'{a}ticas, Universidad Michoacana de San Nicol\'as
de Hidalgo, Ciudad Universitaria, CP 58040, Morelia, Michoac\'{a}n, M\'{e}xico}
\affiliation[d]{Centro de Ci\^encias Naturais e Humanas, Universidade Federal do ABC, 09210-580, Santo Andr\'e, SP, Brazil}
\affiliation[e]{Centro de Matem\'atica, Computa\c c\~ao e Cogni\c c\~ao, Universidade Federal do ABC, 09210-580, Santo Andr\'e, SP, Brazil}

\emailAdd{gabriel@fis.unam.mx}
\emailAdd{malagon@fis.unam.mx}
\emailAdd{aherrera@ifuap.buap.mx}
\emailAdd{andre.kuerten@ufabc.edu.br}
\emailAdd{roldao.rocha@ufabc.edu.br}

\abstract{Within the braneworld paradigm the tachyonic scalar field has been used to generate models that attempt to solve some of the open problems that physics faces nowadays, both in cosmology and high energy physics as well. When these field configurations are produced by the interplay of higher dimensional warped gravity with some matter content, braneworld models must prove to be {\it stable} under the whole set of small fluctuations of the gravitational and matter fields background, among other consistency tests.
Here we present a complete proof of the {\it stability} under scalar perturbations of tachyonic thick braneworlds with an embedded maximally symmetric 4D space-time, revealing its physical consistency. 
This family of models contains a recently reported tachyonic de Sitter thick braneworld which possesses a series of appealing properties. These features encompass complete regularity, asymptotic  flatness (instead of being asymptotically dS or AdS) even when it contains a negative bulk cosmological constant, a relevant 3-brane with dS metric which naturally arises from the full set of field equations of the 5D background (it is not imposed), qualitatively describing the inflationary epochs of our Universe, and a graviton spectrum with a single zero mode bound state that accounts for the 4D graviton localised on the brane and is separated from the continuum of Kaluza-Klein massive graviton excitations by a mass gap. The presence of this mass gap in the graviton spectrum makes the extra-dimensional corrections to Newton's law decay exponentially. Gauge vector fields with a single massless bound state in its mass spectrum are also localised on this braneworld model a fact that allows us to recover the Coulomb's law of our 4D world.
All these properties of the above referred tachyonic braneworld together with the positive stability analysis provided in this work, constitute a firm step towards the construction of realistic cosmological models within the braneworld paradigm.}

\arxivnumber{1508.03867}

\begin{document}
\maketitle
\flushbottom

\section{Introduction}
\label{sec:intro}

The braneworld model paradigm has been very useful to address several open problems in modern physics that cover a wide range of phenomena, from cosmology  \cite{RubakovShaposhnikov,RDW,Langlois}, astrophysics \cite{bwstars,bwstars1,bwstars2,bwstars3,bwstars4,bwDM,cmb1,cmb2,cmb3}, gravity \cite{RS2,LR,Mannheim,maartenskoyama}, high energy physics \cite{Antoniadis,ADD,AADD,gogber,RS1} and even low energy physics \cite{LMTPP,LMTP,LMTPP3}, leading to a plausible reformulation or solution of some of them (see \cite{Langlois,maartenskoyama,reviewRubakov,thickreview} for relevant reviews on this subject). Within this framework, braneworlds are required to render the 4D physics of our Universe at certain physical limit with small corrections to its well established 4D physical laws in such a way that these corrections do not contradict current experimental and/or observational data. One more fundamental test that braneworld models must pass in order to be considered consistent from the physical point of view relies on their stability under small fluctuations of all the background fields (metric and matter fields) that give rise to this class of field configurations. 
Metric perturbations are classified according to the transformations associated with the symmetry group of a given maximally symmetric 4D spacetime (Minkowski or (anti-)de Sitter) into tensor, scalar and vector modes. These modes evolve independently at linear order causing the decoupling of their dynamical equations. Checking for this kind of stability is a non-trivial task that is not simple to afford for models possessing a complicated structure in terms of their field content. In fact, their perturbed Einstein and field equations are extremely non-linear, as in the case of the above mentioned expanding tachyonic thick braneworld. 

As a matter of fact, it is difficult to construct satisfactory, consistent and stable tachyonic braneworld models that yield the phenomenology of our 4D world in certain physical limit. For instance, when considering a braneworld modeled by warped gravity in conjunction with a tachyonic scalar field \cite{Bazeiaetal,koleykar}, the corresponding field equations are highly non-linear and do not allow one to easily find thick brane solutions with a decaying warp factor which leads to the localisation of 4D gravity and other matter fields. Moreover, the analytical study of its stability becomes rather complicated due to the involved structure of the tachyonic scalar field Lagrangian. In this case, stability studies quite often are restricted to a linearised version of the complete model \cite{Bazeiaetal,FKS,German} since the exhaustive analysis requires much more effort to be accomplished.

Moreover, the tachyonic braneworld models constructed so far do not always render the physical laws of our 4D world in a plausible limit. In particular, when attempting to solve the hierarchy gauge problem simultaneously recovering the Newton's law of gravity in the weak field limit, tachyonic field configurations with a single brane turn out to be either complex or fail to localise 4D gravity on the 3-brane. An explicit example of this situation can be found, for instance, in \cite{koleykar,palkar,GRG2014}. Thus, one must perform a complete analytical study of these kind of tests when looking at the physical consistency of a given braneworld model.

A concrete family of expanding (with a 3-brane possessing de Sitter symmetry) tachyonic braneworld models has been recently reported in the literature \cite{German}. This feature may be used to model the inflationary stages of the evolution of our Universe and is compatible with the Lambda Cold Dark Matter scenario that we observe nowadays. In that work the stability analysis of this tachyonic scalar-tensor field configuration under the transverse traceless sector of tensor perturbations was performed in a quite straightforward way. Moreover, the corrections to Newton's law were analytically computed (since an explicit expression for the KK massive modes was obtained) and shown to decay exponentially. However, the stability of this field system under the sector of scalar fluctuations was considered only for the small gradient limit or slow roll approximation of the tachyonic field. This scalar perturbation sector is relevant for the stability of the braneworld since it is partially generated by a tachyonic scalar field. However, for this particular tachyonic field braneworld configuration the squared gradient of the tachyon field is a not so small quantity at the 3-brane position: $\left.(\nabla T)^2\right\vert_{w=0}=\frac{1}{4}$. This points to the need of performing a more careful analysis regarding the stability of such a braneworld model under the scalar sector of small fluctuations.

In this work we fulfill such a missing gap providing an exhaustive analytic study of these scalar perturbations and show that the aforementioned tachyonic braneworld model (with a negative bulk cosmological constant and a positive definite self-interacting tachyonic potential) is {\it stable} under them. Moreover, we shall prove that such braneworld is stable for any maximally symmetric 4D spacetime, either (anti)-de Sitter or Minkowski, embedded in  the considered 5D manifold.

It is worth mentioning as well  that the zero modes of some of the Standard Model matter fields were also shown to be localised on the expanding 3-brane of the above referred tachyonic braneworld, proving that the model is physically viable from the high energy physics point of view. In particular, a gauge vector field and certain phases of a massive scalar field were shown to be localised in \cite{scalvect,scalvect1}, while the localisation of fermions was treated in \cite{fermions}, where the Coulomb's law was recovered as well in the non-relativistic limit of the Yukawa interaction of localised on the brane fermionic and gauge bosonic fields in this inflationary tachyonic thick braneworld scenario. The localisation of these Standard Model matter fields also opens a window for performing high energy physics phenomenology within the framework of this tachyonic braneworld.

On the other hand, this kind of expanding thick de Sitter braneworld models is interesting from a cosmological perspective as well since they qualitatively describes both the inflationary period of the early Universe and the accelerating expansion that we observe nowadays (the second derivative of the scale factor with respect to time is positive). In particular, starting from the fact that the cosmic inflation theory is in good agreement with the temperature fluctuation properties observed in the Cosmic Microwave Background Radiation, and that inflation likely took place at very high temperatures (this study involves several assumptions related to the relevant physical phenomena that take place at such high energies), cosmologists have performed several attempts to construct consistent inflationary models within the framework of string theory, supergravity and now, the braneworld paradigm \cite{popeetal,burgess,SC,DGP}.

The paper is organized as follows: In section II we first show that the braneworld models generated by gravity and a canonical scalar field are stable under scalar fluctuations. We further consider a tachyonic thick braneworld supported by a maximally symmetric 3-brane and prove that it is stable under the same sector of small perturbations. In section III summarize our results and discuss about their physical implications.

\section{Stability of canonical and tachyonic scalar field braneworlds}

In order to better understand the analytical path followed in this work, we shall first review the framework regarding a braneworld model generated by 5D gravity minimally coupled to a canonical scalar field is stable under the scalar sector of perturbations of the field system \cite{Bardeen,Giovannini,KKS}. We therefore shall continue to consider the stability of a regular tachyonic braneworld under this sector of scalar fluctuations and to show that despite the technical difficulties encountered in this process, the model is stable as well under the scalar sector of fluctuations.

\subsection{The canonical scalar field braneworld}

We shall start by considering the following braneworld 5D action generated by a canonical scalar field minimally coupled to gravity  
\begin{eqnarray} 
S=\int d^{5}x\sqrt{-g}\left[ \frac{R}{2}-\frac{1}{2}g^{AB}\partial
_{A}\varphi \partial _{B}\varphi -V(\varphi )\right] ,
\label{action1}
\end{eqnarray} %
with a space-time metric given by%
\begin{eqnarray}
ds_{5}^{2}=a(y)^{2}\gamma _{\mu \nu }dx^{\mu }dx^{\nu
}+dy^{2}=a(w)^{2}\left( \gamma _{\mu \nu }dx^{\mu }dx^{\nu }+dw^{2}\right) ,
\label{metric}
\end{eqnarray}%
where $a(w)$ is the warp factor depending exclusively on the fifth dimension and $\gamma _{\mu \nu}$ represents a 4D space-time of
Minkowski $\mathcal{M}_{4}$, de Sitter $dS_{4}$ or anti de Sitter $AdS_{4}$ form. 
Capital Latin indices, $A,B,$ run from $0$ to $5$, and Greek ones, $\mu,\nu,\alpha,$ label the 4D 3-brane dimensions.
Thus, the field equations for this model are%
\begin{eqnarray}
\varphi ^{\prime\prime} +3\mathcal{H}\varphi ^\prime -a^{2}\frac{%
\partial V}{\partial \varphi }&=&0\,,\\
3\mathcal{H}^{\prime} +3\mathcal{H}^{2}-3K&=&-\frac{\varphi ^{{\prime }2}}{%
2}-a^{2}V,\\
6\mathcal{H}^{2}-6K&=&\frac{\varphi ^{{\prime }2}}{2}-a^{2}V,\\
\mathcal{H}&=&\frac{a^{\prime}}{a},
\end{eqnarray}%
where primes stand for derivatives with respect to the extra dimension, $K$ is the 
4D spatial curvature with $K=0,+1,-1$ for $\mathcal{M}%
_{4}$, $dS_{4}$ and $AdS_{4}$, respectively.\ As an example, their
respective solutions, for the case $K=0$, are given by the following expressions:%
\begin{eqnarray}
a(w)&=&\frac{1}{\sqrt{b^{2}w^{2}+1}},\\
\varphi(w)&=&\sqrt{6}\arctan (bw),\\
V(\varphi)&=&3b^{2}\left[1-5\sin ^{2}\left(\frac{\varphi }{\sqrt{6}}\right)\right] ,
\label{Vphi}
\end{eqnarray}
hence constituting the thick braneworld solution modeled by gravity and the canonical scalar field constructed in the second paper of \cite{Giovannini}. Here is important to note that by determining the desired behaviour of the geometry, the warp factor in this case, the scalar field dynamics is completely fixed and hence its potential.\footnote{One could proceed the other way around when solving this field equations, fixing the scalar field dynamics and solving for the warp factor. However, in this case there is no guarantee that 4D gravity will be localized.} 

\subsection{Scalar sector of fluctuations for the canonical scalar field braneworld}

For the scalar sector of metric fluctuations with arbitrary constant spatial curvature, we shall consider the following perturbed metric \cite{Bardeen,Giovannini,KKS}
\begin{eqnarray} 
ds_{5}^{2}=a^2(w)\left[ \left( 1+2\phi \right) dw^{2}+\left( 1+2\psi
\right) \gamma _{\mu \nu }dx^{\mu }dx^{\nu }\right]\,.
\end{eqnarray} %
Hence, the respective perturbation equations read \cite{KKS} (below we explicitly represent each component) \footnote{For some basic geometrical relations of the fluctuated metric see Appendix A.}
\begin{eqnarray} 
_{z}^{z}&:& \qquad   3\gamma ^{\mu \nu }\overset{\gamma }{\nabla }_{\mu }\overset{\gamma 
}{\nabla }_{\nu }\psi +12\mathcal{H}\psi ^\prime -12\mathcal{H}^{2}\phi
+12K\psi =\varphi _{0}^{\prime} \delta \varphi ^\prime -\phi \varphi
_{0}^{{\prime }2}-a^{2}\frac{\partial V}{\partial \varphi }, \\
%
_{\mu }^{z}&:& \qquad   -3\partial _{\mu }\psi ^\prime +3\mathcal{H}\partial _{\mu
}\phi =\varphi _{0}^{\prime} \partial _{\mu }\delta \varphi , \\
%
_{\nu }^{\mu }&:&\qquad  \left[ 3\psi ^{\prime\prime} -6\mathcal{H}^{{\prime
}}\phi -3\mathcal{H}\phi ^\prime +9\mathcal{H}\psi ^\prime -6%
\mathcal{H}^{2}\phi +\gamma ^{\alpha \beta }\overset{\gamma }{\nabla }%
_{\alpha }\overset{\gamma }{\nabla }_{\beta }\left( \phi +2\psi \right)
+6K\psi \right] \delta _{\nu }^{\mu } \nonumber \\
&& \ \ \ \ \ \ -\gamma ^{\mu \alpha }\overset{\gamma }{\nabla }_{\alpha }\overset{\gamma }%
{\nabla }_{\nu }\left( \phi +2\psi \right) =\left( -\varphi _{0}^{{\prime
}}\delta \varphi ^\prime +\phi \varphi _{0}^{{\prime }2}-a^{2}\frac{%
\partial V}{\partial \varphi }\right) \delta _{\nu }^{\mu }\,,
\end{eqnarray}
where $\overset{\gamma }{\nabla }_{\alpha }$ denotes covariant derivatives with respect to the 4D metric $g_{\mu\nu}$ and we have decomposed the scalar field into its background field configuration value plus a small perturbation: $\varphi =\varphi _{0}+ \delta \varphi$. From the above equations it follows that $\phi =-2\psi$ when $\mu \neq \nu$. Moreover, the components $_{\mu }^{z}$ imply that\footnote{Here some integration functions of the extra dimension were set to zero since, in principle, $\phi = -2\psi + c_{\nu}(w)x^{\nu} + d(w)$ as quoted in \cite{Giovannini}. Notwithstanding, we are interested in a gauge in which $\phi$ and $\psi$ represent wave functions of a graviscalar "particle". This particle freely moves in 4D, therefore, the 4D part of these functions must possess the form $e^{k_{\nu}x^{\nu}}$, where $k_{\nu}$ is a wave 4-vector. However $c_{\nu}(w)x^{\nu} + d(w)$ does not have this form and, therefore, $c_{\nu}(w)$ and $d(w)$ must vanish. In other words, $c_{\nu}(w)$ and $d(w)$ do not represent propagating degrees of freedom and within the problem under consideration one should be interested just in particles/waves that freely propagate in 4D.} 
\begin{eqnarray}
\delta \varphi =-\frac{3}{\varphi _{0}^{\prime} }%
\left( \psi ^\prime + 2\mathcal{H}\psi \right) .  \label{fffm}
\end{eqnarray}%
Then by combining the previous relations we obtain
\begin{eqnarray} 
\psi ^{\prime\prime} +\left[ 3\mathcal{H}-2\frac{\varphi
_{0}^{\prime\prime} }{\varphi _{0}^{\prime} }\right] \psi
^{\prime} +\left[ 4\mathcal{H}^{\prime} -4\mathcal{H}\frac{\varphi
_{0}^{\prime\prime} }{\varphi _{0}^{\prime} }-4K\right] \psi
=\left(\gamma ^{\mu \nu }\overset{\gamma }{\nabla }_{\mu }\overset{\gamma }{%
\nabla }_{\nu }-2K\right)\psi \,.
\text{ }
\label{tfs}
\end{eqnarray} %
Now, it is convenient to propose the following \emph{ansatz}%
\begin{eqnarray}
\psi (w,x^{\mu })=\frac{\varphi _{0}^{\prime} (w)}{a(w)^{3/2}}F(w,x^{\mu
}),  
\end{eqnarray}%
in order to rewrite (\ref{tfs}) as a Schr\"odinger-like equation%
\begin{eqnarray}
-F^{\prime\prime} +\mathcal{V}(w)F=-\left(\gamma ^{\mu \nu }\overset{\gamma }{\nabla }_{\mu }\overset{\gamma }{%
\nabla }_{\nu }-2K\right)F,  \label{els}
\end{eqnarray}%
where the potential $\mathcal{V}(w)$\ is given by the relation%
\begin{eqnarray} 
\mathcal{V}=-\frac{5}{2}\mathcal{H}^{\prime} +\frac{9}{4}\mathcal{H}^{2}+%
\mathcal{H}\frac{\varphi _{0}^{\prime\prime} }{\varphi _{0}^{{\prime
}}}-\frac{\varphi _{0}^{{{\prime \prime \prime }}}}{\varphi _{0}^{{\prime
}}}+2\left( \frac{\varphi _{0}^{\prime\prime} }{\varphi
_{0}^{\prime} }\right) ^{2}-4K.
\end{eqnarray} 

Giovannini analyzed the left hand side of Eq. (\ref{els}) after separating variables \cite{Giovannini}, for
the case $K=0$, and defined the following superpotential operators\footnote{The term superpotential is borrowed from supersymmetric quantum mechanics.}
\begin{eqnarray}
\Lambda ^{\dagger }=\left[ -\frac{\partial }{\partial w}+\mathcal{J}\right] 
\text{ \ \ and \ \ }\Lambda =\left[ \frac{\partial }{\partial w}+\mathcal{J}%
\right] ,  \label{ogio}
\end{eqnarray}%
such that 
\begin{eqnarray}
\Lambda ^{\dagger }\Lambda F=-F^{\prime\prime} +(\mathcal{J}^{2}-%
\mathcal{J}^{\prime} )F=m_{F}^{2}F,\qquad \text{ \ \ }K=0\,.  \label{ssoe}
\end{eqnarray}%
The relation between the potential $\mathcal{V}(w)$ and the
superpotential $\mathcal{J}(w)$ must satisfy 
\begin{eqnarray}
\mathcal{V}=\mathcal{J}^{2}-\mathcal{J}^{\prime} \, \label{pgio}
\end{eqnarray}%
and above we have used the definition of the 4D mass\footnote{Note that we use the signature $(-,+,+,+,+)$ throughout the paper, which is the opposite compared to the one used in \cite{Giovannini}. By the way, it should be pointed out as well that there is a misprint in the sign of the mass definition given in \cite{German}.} 
\begin{equation}
\overset{\eta }{%
\nabla^{\mu }}\overset{\eta }{\nabla }_{\mu }\xi = m_{\xi }^{2}\xi 
\label{mdSK0}
\end{equation} 
for a scalar field $\xi $ in a Minkowski space-time (when $K=0$). It turns out that 
the hermiticity and positive definite character of the left hand side of Eq. (\ref{ssoe}) guarantee a positive spectrum of scalar perturbations, evidencing the lack of tachyonic unstable modes with $m^2 < 0$ and ensuring 
the stability of the braneworld configuration under the whole sector of small scalar fluctuations.

Remarkably, these results can be straightforwardly generalized to space-times with constant 
curvature, i.e. they are valid for either $M_4$, $dS_4$ or $AdS_4$, and the definition
of the 4D mass should be modified correspondingly \cite{mdS,mdS1,mdS2,wang}, in 
particular, for the de Sitter space, the 4D mass operator is defined as 
follows: 
\begin{equation}
\overset{\eta }{%
\nabla^{\mu}}\overset{\eta }{\nabla }_{\mu }-2H^2 \equiv m^{2},
\label{mdS}
\end{equation}
where $H$ labels the Hubble parameter and is related to the spatial curvature
as $K=H^2$. This relation shows that the expansion of space-time actually
contributes to the mass of a body located in its gravitational field.

Since Eq. (\ref{ssoe}) asserts that $m_{F}^{2}>0$, hence the canonical
scalar braneworld configuration is stable if $\mathcal{J}$ does exist and fulfills
the relation (\ref{pgio}). Now, by rewriting the superpotential $\mathcal{J}$ as%
\begin{eqnarray}
\mathcal{J=-}\frac{\mathcal{Z}^{\prime} }{\mathcal{Z}}\text{ \ \ with \ \ 
}\mathcal{Z}^{-1}=\frac{1}{\mathcal{H}}a^{3/2}\varphi _{0}^{\prime} ,
\label{JZ}
\end{eqnarray}%
we obtain the following expression for the potential 
\begin{eqnarray}
\text{\ \ }\mathcal{V}=\frac{\mathcal{Z}^{\prime\prime} }{\mathcal{Z}}=-%
\frac{5}{2}\mathcal{H}^{\prime} +\frac{9}{4}\mathcal{H}^{2}+\mathcal{H}%
\frac{\varphi _{0}^{\prime\prime} }{\varphi _{0}^{\prime} }-\frac{%
\varphi _{0}^{{{\prime \prime \prime }}}}{\varphi _{0}^{\prime} }%
+2\left( \frac{\varphi _{0}^{\prime\prime} }{\varphi _{0}^{\prime} %
}\right) ^{2}, \qquad K=0, \label{VZ}
\end{eqnarray}%
rendering a further quite simplified expression that will be useful when analyzing 
the stability of the expanding tachyonic braneworld model under consideration.

\subsection{The expanding tachyonic braneworld model}

Now, we shall briefly review the expanding tachyonic braneworld. We shall start from a string theory perspective of the derivation of the effective action from which our model was inspired and finish with an analysis of the physical properties of a concrete solution for this field configuration within the braneworld realm. 

The action for the model we shall consider reads 
\begin{eqnarray}
S=\int d^{5}x\sqrt{-g}\left[ \frac{R}{2\kappa _{5}^{2}}-\Lambda _{5}-V(T)%
\sqrt{1+g^{AB}\partial _{A}T\partial _{B}T}\right]\,,  \label{ta}
\end{eqnarray}
and describes gravity with a bulk cosmological constant $\Lambda_5$ minimally coupled to a real tachyonic scalar field, where $V(T)$ denotes its self-interaction potential, $\kappa_5$ stands for the 5D gravitational coupling constant and again $A,B=0,1,2,3,5.$ 

The form of the tachyonic effective action in Eq. (\ref{ta}) was originally proposed in \cite{Sen0} as a supersymmetric generalization of the Dirac-Born-Infeld (DBI) action describing the dynamics of light modes (tachyonic and massless) on the world-volume of a non-BPS D-brane within the framework of type II string theory in Minkowski spacetime (for a progressive and illustrative construction of this action see as well \cite{garousi,bergshoeff,kluson}). It turns out that in the context of string theory, D-branes can give rise to both stable BPS states and unstable objects such as brane-antibrane configurations and non-BPS D-branes (for a review see \cite{senWSL}). In particular, non-BPS branes in Type II string theories are unstable and can decay to stable D-branes: a non-BPS Dp-brane in the theory (either type IIA or type IIB) will condense to a BPS D(p-1)-brane. Moreover, the non-BPS Dp-branes in these theories are related to BPS D(p+1)-brane-antibrane configurations by condensation of the tachyon living on this brane-antibrane pair. Thus, the tachyonic effective field theory describing the dynamics of a non-BPS D-brane in string theory possesses a BPS D-brane. By studying the world-volume theory of massless modes on this BPS D-brane, it was shown that the world volume action has precisely the Dirac-Born-Infeld form without any higher derivative corrections \cite{sen4}. In this tachyonic effective field theory the self-interaction potental $V(T)$ is symmetric under $T\to -T$, has a maximum at $T = 0$, and possesses its minimum at $T = \pm\infty$ where it vanishes.\footnote{The tachyon field we consider in the action (\ref{ta}) is real in contrast to the fact that the original string theory effective action involves a complex tachyon field that lives in the brane-antibrane configuration.}

The tachyonic effective action given by Eq. (\ref{ta}) has found several applications within braneworld cosmology \cite{mazumdar} and string cosmology \cite{sen,sen1,sen2,sen3,gibbons,chodd,chodd1,padmanabhan}; in particular, some aspects of canonical quantization of this field theory coupled to gravity were studied in \cite{sen3}, where the possibility of using the tachyonic scalar field as the definition of time in quantum cosmology was explored. Moreover, by considering this action as a scalar-tensor model, solar system constraints were imposed on its parameters in \cite{devi}. Recent studies of tachyon inflation within the $N$-formalism, which takes a prescription for the small Hubble flow slow-roll parameter $\epsilon_1$ as a function of a large number of $N$ $e$-folds, have lead to an analysis of observables in the light of the Planck 2015 data and show the viability of some of this class of models \cite{nan}.

The whole set of field equations are derived from (\ref{ta}) when $\delta S=0$, and read
\begin{eqnarray} 
\square T-\frac{\nabla _{C}\nabla _{D}T\nabla ^{C}T\nabla ^{D}T}{1+\nabla
_{A}T\nabla ^{A}T}&=&\frac{1}{V}\partial _{T}V,\\
R_{AB}-\frac{1}{2}Rg_{AB}&=&\kappa _{5}^{2}\left( T_{AB}^{\mathcal{B}}-\Lambda
_{5}g_{AB}\right) ,
\end{eqnarray} %
where $T_{AB}^{\mathcal{B}}$ denotes the bulk stress energy tensor given by 
\begin{eqnarray} 
T_{AB}^{\mathcal{B}}=-g_{AB}V(T)\sqrt{1+\left( \nabla T\right) ^{2}}+\frac{%
V(T)}{\sqrt{1+\left( \nabla T\right) ^{2}}}\partial _{A}T\partial _{B}T.
\end{eqnarray} 

Hereupon we shall consider the following metric \emph{ansatz}\footnote{In what follows the warp factor 
of the expression (\ref{metric}) will be denoted as $a(w)=e^{2f(w)}$.}
\begin{equation}
ds^2 = e^{2f(w)} \left[- d t^2 + a^2(t) \left( dx^2 + dy^2 + dz^{2}\right) + dw^2\right].
\label{ansatzconf}
\end{equation}
where $a(t)$ is the scale factor of the 3-brane.

Therefore, the aforementioned field equations become
\begin{eqnarray}
T^{\prime\prime}-f^{\prime}T^{\prime}+4f^{\prime}T^{\prime}(1+e^{-2 f}T^{\prime2})&=&(e^{2 f}+T^{\prime2})\frac{\partial_{T}V(T) }{V(T)},
\label{Tz}\\
f^{\prime\prime} - f^{\prime2} + \frac{\ddot a}{a} &=& - \kappa_5^2\frac{V(T)T^{\prime2}}{3\sqrt{1 + e^{-2 f}T^{\prime2}}},
\label{Ee1z} \\
f^{\prime2} + \frac{\kappa_5^2\,\Lambda_5}{6}e^{2 f} - \frac{1}{2} \left(\frac{\ddot a}{a}+\frac{\dot a^2}{a^2} \right) &=& - \kappa_5^2\frac{e^{2 f}V(T)}{6\sqrt{1
+ e^{-2 f}T^{\prime2}}}, 
\label{Ee2z}
\end{eqnarray}
where the primes stand again for derivatives with respect to $w$, while the dots do it for time derivatives.

A concrete solution to this set of equations was found in \cite{German} and reads
\begin{eqnarray}
a(t)&=&e^{H\,t}\, 
\label{scalefactor}\\
f(w)&=&-\frac{1}{2}\ln\left[\frac{\cosh\left[\,H\,(2w+c)\right]}{s}\right], \label{fw}\\
T(w) &=& \pm\sqrt{\frac{-3}{2\,\kappa_5^2\,\Lambda_5}}\
\mbox{arctanh}\left[\frac{\sinh\left[\frac{H\,\left(2w+c\right)}{2}\right]}
{\sqrt{\cosh\left[\,H\,(2w+c)\right]}}\right], 
\label{Tw}\\
V(T) &=& - \Lambda_5\ \mbox{sech}\left(\sqrt{-\frac{2}{3}\kappa_5^2\,\Lambda_5}\ T\right)
\sqrt{6\ \mbox{sech}^2\left(\sqrt{-\frac{2}{3}\kappa_5^2\,\Lambda_5}\ T\right)-1}\nonumber \\
&=&
 - \frac{\Lambda_5}{\sqrt{2}}\ \sqrt{(1+\mbox{sech}\left[\,H\,(2w+c)\right])({2+3\
\mbox{sech}\left[\,H\,(2w+c)\right])}}\,, \label{VT}
\end{eqnarray}
where $H$, $c$ and $s>0$ are constants, and 
\begin{equation}
s=-\frac{6H^2}{\kappa_5^2\,\Lambda_5}\,, 
\label{s}
\end{equation}
Moreover, the bulk cosmological constant must be negative definite $\Lambda_5<0$ for consistency. 

The warp factor possesses a decaying behavior and asymptotically vanishes, whereas the tachyon scalar field is real and has a kink/antikink profile. It is worth noticing that in this model it is possible to explicitly express the self-interacting tachyon potential in terms of the tachyonic scalar field $V(T)$. This potential has a maximum at the position of the brane and is positive definite when $\Lambda_5<0$ as it can easily be seen from (\ref{VT}). 

We finally would like to analyze the 5D curvature scalar of this field configuration:
\begin{equation}
R=-\frac{14}{3}\kappa_5^2\,\Lambda_5\,\mbox{sech}\left[\,H\,(2w+c)\right]. 
\label{R5}
\end{equation}
This 5D invariant is positive definite and asymptotically vanishes, yielding an asymptotically 5D Minkowski space-time. The presence of a negative bulk cosmological constant $\Lambda_5<0$ leads to asymptotically $AdS_5$ space-times in the absence of matter. However, by looking at the equation for the action (\ref{ta}) it is evident that the overall effective cosmological constant of the 5D space-time has two relevant contributions: the bulk cosmological constant itself $\Lambda_5$ and the self--interaction potential $V(T)$, which asymptotically adopts the value $-\Lambda_5$, rendering and asymptotically flat 5D space-time. 

It is quite straightforward to show that in the limit when $H^2\to 0$, the whole braneworld field configuration blows up unless we take this limit when $\Lambda_5\to 0$ simultaneously in such a way that $s$ remains finite (see \cite{German} for further details). This procedure renders a linear tachyonic field with a vanishing self-interaction potential, and constant scale and warp factors that lead to a 5D Minkowski spacetime after a coordinate rescaling. Thus, this is a non-perturbative solution in the sense that it does not posses a perturbative flat 4D spacetime limit.

Thus, the solution (\ref{scalefactor})-(\ref{s}) constitutes a non-trivial, regular, asymptotically flat 5D tachyonic braneworld configuration (it interpolates between two 5D Minkowski spacetimes) with an expanding induced metric on the brane described by a de Sitter 4D cosmological background, supported by a bulk kink tachyonic scalar field and a negative cosmological constant.

The actions for the canonical scalar field and the tachyonic scalar field are quite different and possess a rather different structure when compared to each other; however, it is worth noticing that in the slow roll approximation, i.e. when $\nabla T << 1,$ the action (\ref{ta}) transforms into (\ref{action1}) under the identifications $\varphi=\varphi(T)$ and $\partial_A\varphi=\sqrt{V(T)}\partial_A T.$ In this limit, the self-interaction potentials (\ref{Vphi}) and (\ref{VT}) are related through the following expression 
\begin{eqnarray}
V(\varphi)=3b^{2}\left[1-5\sin ^{2}\left(\frac{2}{3}V(T)^{3/2}\right)\right] .
\end{eqnarray}

\subsection{Scalar sector of perturbations for the tachyonic thick braneworld}

The stability analysis of this tachyonic braneworld model under tensor fluctuations was completely performed in \cite{German}, whereas its stability under scalar perturbations was presented for the restricted case of slow roll approximation, where the kinetic term of the tachyonic action is small $g^{AB}\partial _{A}T\partial _{B}T\ll 1$. Here we shall study  the most general case with a maximally symmetric metric $\gamma _{\mu\nu}$ without any approximation for the tachyonic action.

The respective fluctuations of this tachyonic braneworld model under scalar perturbations are provided by the following fluctuated relations (see Appendix B for details)
\begin{eqnarray}
\delta \square T-\delta \left[ \frac{\nabla _{C}\nabla _{D}T\nabla
^{C}T\nabla ^{D}T}{1+\nabla _{A}T\nabla ^{A}T}\right] &=&\delta \left[ \frac{1}{V}\partial _{T}V\right] ,  \label{f1}\\
\delta R_{A}^{B}-\frac{1}{2}\delta _{A}^{B}\delta R&=&\kappa _{5}^{2}\delta (T^{\mathcal{B}}){}_{A}^{B},  \label{f2}
\end{eqnarray} 
where the perturbed energy--momentum tensor of the tachyonic fields reads
\begin{eqnarray}
\delta T^{\mathcal{B}}{}_{A}^{B} &=&-\left[ \sqrt{1+e^{-2f}T_{0}^{^{\prime
}2}}\left( \partial _{T_{0}}V\right) \delta T+\frac{e^{-2f}T_{0}^{{\prime
}}V\left( \delta T^{\prime} -\phi T_{0}^{\prime} \right) }{\sqrt{%
1+e^{-2f}T_{0}^{{\prime }2}}}\right] \delta _{A}^{B}  \notag \\
&&+e^{-2f}T_{0}^{{\prime }2}\left[ \frac{\left( \partial _{T_{0}}V\right)
\delta T}{\sqrt{1+e^{-2f}T_{0}^{{\prime }2}}}+\frac{e^{-2f}T_{0}^{{\prime
}}V\left( \phi T_{0}^{\prime} -\delta T^{\prime} \right) }{\left(
1+e^{-2f}T_{0}^{{\prime }2}\right) ^{3/2}}\right] \delta _{A}^{w}\delta
_{w}^{B}\nonumber\\&&+\frac{T_{0}^{\prime} V}{\sqrt{1+e^{-2f}T_{0}^{{\prime }2}}}\left[
\delta _{A}^{w}\mathring{g}^{BC}\partial _{C}\delta T+e^{-2f}\left(
\delta _{w}^{B}\partial _{A}\delta T-2\delta _{A}^{w}\delta _{w}^{B}\phi
T_{0}^{\prime} \right) \right] ,  \label{f3}
\end{eqnarray}
and we used $T=T_0+\delta T$. Eqs. (\ref{f2}) and (\ref{f3}) imply that 
\begin{eqnarray}
3\gamma ^{\mu \nu }\overset{\gamma }{\nabla }_{\mu }\overset{\gamma }{\nabla 
}_{\nu }\psi +12f^{\prime} \psi ^\prime -12f^{{\prime }2}\phi
+12K\psi &=&\kappa _{5}^{2}e^{2f}\delta T^{\mathcal{B}}{}_{w}^{w},  \label{e1}\\
-3\partial _{\mu }\psi ^\prime +3f^{\prime} \partial _{\mu }\phi
&=&\kappa _{5}^{2}e^{2f}\delta T^{\mathcal{B}}{}_{\mu }^{w},  \label{e2}
\end{eqnarray}
and moreover
\begin{eqnarray}
&&\left[ 3\psi ^{\prime\prime} -6f^{\prime\prime} \phi
-3f^{\prime} \phi ^\prime +9f^{\prime} \psi ^{{\prime
}}-6f^{{\prime }2}\phi +\gamma ^{\alpha \beta }\overset{\gamma }{\nabla }%
_{\alpha }\overset{\gamma }{\nabla }_{\beta }\left( \phi +2\psi \right)
+6K\psi \right] \delta _{\nu }^{\mu }  \notag \\
&&-\gamma ^{\mu \alpha }\overset{\gamma }{\nabla }_{\alpha }\overset{\gamma }%
{\nabla }_{\nu }\left( \phi +2\psi \right) =\kappa _{5}^{2}e^{2f}\delta T^{%
\mathcal{B}}{}_{\nu }^{\mu }\,. \label{e3}
\end{eqnarray}
The variations of the energy-momentum tensor thus can be derived, rendering the following relations:
\begin{eqnarray}
\label{Tww}
\delta T^{\mathcal{B}}{}_{w}^{w} &=&-\left[ \sqrt{1+e^{-2f}T_{0}^{^{\prime
}2}}\left( \partial _{T_{0}}V\right) \delta T+\frac{e^{-2f}T_{0}^{{\prime
}}V\left( \delta T^{\prime} -\phi T_{0}^{\prime} \right) }{\sqrt{%
1+e^{-2f}T_{0}^{{\prime }2}}}\right] \\
&&+\frac{e^{-2f}T_{0}^{{\prime }2}}{\sqrt{1+e^{-2f}T_{0}^{{\prime }2}}}%
\left[ \left( \partial _{T_{0}}V\right) \delta T+\frac{e^{-2f}T_{0}^{^{%
\prime }}V\left( \phi T_{0}^{\prime} -\delta T^{\prime} \right) }{%
1+e^{-2f}T_{0}^{{\prime }2}}\right] +\frac{2e^{-2f}T_{0}^{\prime} V}{%
\sqrt{1+e^{-2f}T_{0}^{{\prime }2}}}\left[ \delta T^{\prime} -\phi
T_{0}^{\prime} \right] , \nonumber \\
\label{Tnumu}
\delta T^{\mathcal{B}}{}_{\nu }^{\mu } &=&-\left[ \sqrt{1+e^{-2f}T_{0}^{^{%
\prime }2}}\left( \partial _{T_{0}}V\right) \delta T+\frac{%
e^{-2f}T_{0}^{\prime} V\left( \delta T^{\prime} -\phi T_{0}^{{\prime
}}\right) }{\sqrt{1+e^{-2f}T_{0}^{{\prime }2}}}\right] \delta _{\nu }^{\mu}, \\
\label{Tmuw}
\delta T^{\mathcal{B}}{}_{\mu }^{w} &=&\frac{e^{-2f}T_{0}^{\prime} V}{%
\sqrt{1+e^{-2f}T_{0}^{{\prime }2}}}\partial _{\mu }\delta T.
\end{eqnarray}

Eqs. (\ref{e3}) and (\ref{Tnumu}) imply, when $\mu \neq \nu $, a relation between the
metric fluctuations $\phi $ and $\psi$ provided by  
\begin{eqnarray}
\phi =-2\psi\,.  
\label{a1}
\end{eqnarray}%
Moreover, Eqs. (\ref{e2}) and (\ref{Tmuw}) provide a relation between the field fluctuation 
$\delta T$ and $\psi $ given by
\begin{eqnarray}
\partial _{\mu }\delta T=\mathcal{F}(w)\partial _{\mu }\left( \psi
^{\prime} +2f^{\prime} \psi \right) \Rightarrow \delta T=\mathcal{F}(w)%
\left( \psi ^\prime +2f^{\prime} \psi \right) ,  
\label{a2}
\end{eqnarray}
where $\mathcal{F}(w)$ reads
\begin{eqnarray} 
\mathcal{F}(w)=-\frac{3\sqrt{1+e^{-2f}T_{0}^{{\prime }2}}}{\kappa
_{5}^{2}T_{0}^{\prime} V}\,.
\end{eqnarray} %
This function is precisely the tachyonic version for the expression (\ref{fffm}) corresponding to the 
canonical scalar field. 

Now, by equating the first terms of the right-hand side of Eqs. (\ref{Tww}) and (\ref{Tnumu}) we get
\begin{eqnarray} 
\delta T^{\mathcal{B}}{}_{w}^{w}=\delta{{T}^{\mathcal{B}%
}{}_{\nu }^{\mu }}\!+\!\frac{e^{-2f}T_{0}^{\prime} }{\sqrt{1+e^{-2f}T_{0}^{^{%
\prime }2}}}\left[ T_{0}^{\prime} \left( \partial _{T_{0}}V\right) \delta
T\!-\left(\!\frac{e^{-2f}T_{0}^{{\prime }2}}{1+e^{-2f}T_{0}^{{\prime }2}}\!-\!2\right)V\left( \delta
T^{\prime}\!+\!2\psi T_{0}^{\prime} \right) \right]\!,
\label{eqpsi}
\end{eqnarray} 
for $\mu =\nu$ with no summation in these indices. We further substitute Eqs. (\ref{e1}) and (\ref{e3}), when $\mu =\nu$ without summation under these indices, into Eq. (\ref{eqpsi}) and make use of the expression (\ref{a2}) for $\delta T$ and its first derivative with respect to the extra coordinate $\delta T^\prime$ for obtaining, after some algebraic work, the following relation 
\begin{eqnarray}
&&\left( \mathcal{A}-1\right) \psi ^{\prime\prime} +\left[ f{^{\prime
}}-\mathcal{D}+\mathcal{B}\left( \mathcal{A-}2\right) \right] \psi
{^{\prime }}+2\left[ 2f^{\prime\prime} -2f^{{\prime }2}-f^{{\prime
}}\mathcal{D}+\mathcal{C}\left( \mathcal{A-}2\right)-4K \right] \psi   \notag \\
&=&\left( \gamma ^{\mu \nu }\overset{\gamma }{\nabla }_{\mu }\overset{\gamma 
}{\nabla }_{\nu }-2K\right) \psi \equiv m^2 \psi.  \label{e4}
\end{eqnarray}%
In the last equality we have taken into account the definition of the 4D mass in a maximally symmetric 
space-time with metric $\gamma_{\mu\nu}$ \cite{mdS}, and the functions $\mathcal{A}(w)$, $\mathcal{B}(w)$, $\mathcal{C}(w)$ and $%
\mathcal{D}(w)$ respectively are %
\begin{eqnarray} \label{mata}
\mathcal{A}=\frac{e^{-2f}T_{0}^{{\prime }2}}{1+e^{-2f}T_{0}^{{\prime }2}%
},\text{ \ \ }\mathcal{B}\mathcal{=}\frac{\mathcal{F}^{{\prime }}}{%
\mathcal{F}}+2f^{{\prime }},\text{ \ \ }\mathcal{C}\mathcal{=}\frac{%
\mathcal{F}^{\prime} }{\mathcal{F}}f^{\prime} +f^{{{\prime \prime
}}}+\frac{T_{0}^{\prime} }{\mathcal{F}}, \text{ \ \ }
\mathcal{D}%
\mathcal{=}T_{0}^{\prime} V^{-1}\partial _{T_{0}}V.
\end{eqnarray} 

Finally, by using the relations (\ref{a1}) and (\ref{a2}) in the expression (\ref{f1}) for the perturbed Klein-Gordon equation for the tachyonic scalar field, we find (see Appendix B for details)%
\begin{eqnarray}
&&\left( \mathcal{A-}1\right) \delta T^{\prime\prime}\!+\!\left[\!
f^{\prime} \left( 2\mathcal{A}^{2}\!-\!3\mathcal{A-}3\right)\!+\!2\frac{%
T_{0}^{\prime\prime} }{T_{0}^{\prime} }\mathcal{A}\left(1\!-\!\mathcal{A}%
\right)\!\right] \delta T^{\prime}\! \notag \\ 
&&+\!\left\{\!\frac{e^{2f}}{V}\left[ \partial\,_{T^{2}}^{2}V\!-\!V^{-1}\!\left(
\partial\,_{T}V\right)^{2}\right]\!-\!\frac{4}{\mathcal{F}}T_{0}^{{\prime
}}\!\right\} \delta T 
\!-\!4\left[\left( \mathcal{A}^{2}\!-\!2\mathcal{A}+1\right) T_{0}^{^{\prime
\prime }}\!-\!f^{\prime} \left( \mathcal{A}^{2}-2\mathcal{A}-1\right)
T_{0}^{\prime} \right]\psi   \notag \\
&=&\gamma ^{\mu \nu }\overset{\gamma }{\nabla }_{\mu }\overset{\gamma }{%
\nabla }_{\nu }\delta T.
\label{perttT}
\end{eqnarray}%
Now we need to lead Eq. (\ref{e4}) into a Schr\"{o}dinger-like form. In order to achieve this aim, we perform the following transformation 
\begin{eqnarray} 
\psi =\frac{1}{\alpha ^{1/2}(w)}%
F\left( w,x^{\mu }\right) ,
\end{eqnarray} %
where the auxiliary function $\alpha$, as well as other details worth looking at, are defined in the Appendix C, to where we refer again the reader.
Thus, after some straightforward but tedious algebraic calculations, we obtain the following Schr\"{o}dinger-like equation%
\begin{eqnarray}
-\left(1 - \mathcal{A}\right) F^{\prime\prime} +\mathcal{V}F=
m^2 F,  \label{ets}
\end{eqnarray}%
where the associated potential $\mathcal{V}$ is given by  
\begin{eqnarray} 
\mathcal{V}=\frac{7}{2}f^{\prime\prime} -4f^{{\prime }2}-\frac{1}{2}%
\mathcal{E}_{1}^{\prime} +2\mathcal{E}_{2}-4K-\frac{1}{2\left( \mathcal{A-}%
1\right) }\left[ \frac{1}{2}f^{{\prime }2}+\frac{1}{2}\mathcal{E}%
_{1}^{2}+f^{\prime} \mathcal{E}_{1}-\mathcal{A}^{\prime} \left(
f^{\prime} +\mathcal{E}_{1}\right) \right] .
\label{potl}
\end{eqnarray} 

Compared to Eq. (\ref{els}), the above Schr\"{o}%
dinger-like equation has a weight $g=1-\mathcal{A}$. For an
equation with weight $g$, we can define supersymmetric operators $\Pi
^{\dagger }$ and $\Pi $ as follows 
\begin{eqnarray} 
\Pi ^{\dagger }=\left[ -g^{1/2}\frac{\partial }{\partial w}+\frac{1}{4}\frac{%
g^{\prime} }{g^{1/2}}+\mathcal{J}\right] \text{ \ \ and \ \ }\Pi =\left[
g^{1/2}\frac{\partial }{\partial w}-\frac{1}{4}\frac{g^{\prime} }{g^{1/2}}%
+\mathcal{J}\right]\,.
\end{eqnarray} %
Hence we can rewrite the expression (\ref{ets}) with the aid of these operators
\begin{eqnarray} 
\Pi ^{\dagger }\Pi F=-gF^{\prime\prime} +\left[ \frac{1}{4}g^{{{\prime
\prime }}}-\frac{3}{16}\frac{g^{{\prime }2}}{g}+\mathcal{J}^{2}-g^{1/2}%
\mathcal{J}^{\prime} \right] F = m^2 F, 
\label{PiPit}
\end{eqnarray} %
where the relation between $\mathcal{J}$ and $\mathcal{V}$ is given by 
\begin{eqnarray} 
\mathcal{V=}\frac{1}{4}g^{\prime\prime} -\frac{3}{16}\frac{%
g^{{\prime }2}}{g}+\mathcal{J}^{2}-g^{1/2}\mathcal{J}^{\prime} .
\end{eqnarray} %
Substituting $g$ in terms of $\mathcal{A}$ we have the following expression for the potential%
\begin{eqnarray} 
\mathcal{V}=\mathcal{J}^{2}-\sqrt{1-\mathcal{A}}\mathcal{J}^{\prime} -%
\frac{1}{4}\mathcal{A}^{\prime\prime} -\frac{3}{16}\frac{\mathcal{A}%
^{{\prime }2}}{1-\mathcal{A}},
\end{eqnarray} %
and also for the supersymmetric operators $\Pi ^{\dagger }$ and $\Pi $:
\begin{eqnarray} 
\Pi ^{\dagger }=\left[ -\sqrt{1-\mathcal{A}}\frac{\partial }{\partial w}-%
\frac{1}{4}\frac{\mathcal{A}^{\prime} }{\sqrt{1-\mathcal{A}}}+\mathcal{J}%
\right] \text{ \ and \ }\Pi =\left[ \sqrt{1-\mathcal{A}}\frac{\partial 
}{\partial w}+\frac{1}{4}\frac{\mathcal{A}^{\prime} }{\sqrt{1-\mathcal{A}}%
}+\mathcal{J}\right].\,
\end{eqnarray} 
It is worth to emphasize  that when we consider the limit $\mathcal{A}\rightarrow 0$ ($g\rightarrow 1$),
we recover the operators (\ref{ogio})-(\ref{pgio}) introduced by Giovannini in \cite{Giovannini}:
\begin{eqnarray}
\Pi ^{\dagger } &\rightarrow &\Lambda ^{\dagger }=\left[ -\frac{\partial 
}{\partial w}+\mathcal{J}\right] ,\text{ \ \ }\Pi \rightarrow \Lambda =%
\left[ \frac{\partial }{\partial w}+\mathcal{J}\right] , \\
\mathcal{V} &\rightarrow &\mathcal{J}^{2}-\mathcal{J}^{\prime} .
\end{eqnarray}

However, we still must prove that writing the Schr\"{o}dinger-like equation in terms
of $\Pi ^{\dagger }$ and $\Pi $ implies that 
\begin{eqnarray} 
\overset{\gamma }{\nabla ^{\mu }}\overset{\gamma }{\nabla }_{\mu}-2K=m_{F}^{2}>0,
\end{eqnarray} 
an expression which coincides with the definition of the 4D mass in a maximally symmetric space-time and is posttive definite, stating that there are no states with negative squared masses.

\subsubsection{Auxiliary Sturm-Liouville eigenvalue problem}

In order to achieve  this aim, let us start by considering the equation corresponding to a Sturm-Liouville eigenvalue problem for the field $u(x)$%
\begin{eqnarray} 
-u^{\prime\prime} (x)+q(x)u(x)=\lambda \omega (x)u(x),
\label{SLeqn}
\end{eqnarray} %
where $\lambda $ is the associated eigenvalue and $\omega (x)$ is the weight of 
$\lambda $ and here $^{(^{\prime })}=d/dx$. By making respectively a change of
variable and a specific redefinition of the field%
\begin{eqnarray}
dy &=&\omega (x)^{1/2}dx, \\
v(y) &=&v(y(x))=\omega (x)^{1/4}u(x),
\end{eqnarray}%
the Sturm-Liouville problem can be rewritten as follows
\begin{eqnarray} 
-v(y)_{yy}+\left[ \frac{1}{4\omega }\left( \frac{\omega ^{\prime\prime} %
}{\omega }-\frac{5\omega ^{{\prime }2}}{4\omega ^{2}}\right) +\frac{q}{%
\omega }\right] v(y)=\lambda v(y),
\end{eqnarray} %
where the derivative of the function $v(y(x))$ is given by $v^{\prime} = \omega ^{1/2}v_{y}$, the index $_{y}$ represents the derivative with respect the variable $y$ and we must note that $\omega $ and $q$ are functions of $x$. Thus we must
make the following changes%
\begin{eqnarray} 
d\xi =\omega (w)^{1/2}dw\text{ \ \ and \ \ }\eta (\xi )=\omega (w)^{1/4}F(w),
\end{eqnarray} %
and then we can rewrite Eq. (\ref{ets}) as%
\begin{eqnarray} 
-F(w)^{\prime\prime} +\mathcal{U(}w\mathcal{)}F=\lambda \omega (w)F(w),
\end{eqnarray} %
where $\mathcal{U=V}/g$, $\lambda =\overset{\gamma }{\nabla ^{\mu }}\overset{%
\gamma }{\nabla }_{\mu }-2K$ and $\omega =g^{-1}$. By accomplishing this procedure, we can obtain
for the last expression, %
by further replacing $\omega ^{-1}\mathcal{U=V}$ and $\omega =g^{-1}$ in it, the following result: %
\begin{eqnarray} 
-\eta (\xi )_{\xi \xi }+\left[ \frac{1}{4}\left( -g^{\prime\prime} +%
\frac{3g^{{\prime }2}}{4g}\right) +\mathcal{V}\right] \eta (\xi )=\lambda
\eta (\xi )\,.
\end{eqnarray} %
Now, expliciting $\mathcal{V}$ in terms of $g$ yields
\begin{eqnarray}
-\eta (\xi )_{\xi \xi }+\left[ \mathcal{J(}w\mathcal{)}^{2}-g(w)^{1/2}%
\mathcal{J}^{\prime} (w)\right] \eta (\xi )=\lambda \eta (\xi ).
\label{ecJ}
\end{eqnarray}%
Finally, let us consider the following expression%
\begin{eqnarray} 
-\eta (\xi )_{\xi \xi }+\mathcal{V(\xi )}\eta (\xi )=\lambda \eta (\xi ),
\end{eqnarray} %
where $\mathcal{V}(\mathcal{\xi })$ is the potential written in terms of $%
\mathcal{\xi }$. As we already know the standard pair of operators $\Lambda
^{\dag }(\xi )$ and $\Lambda (\xi )$, which acting on the function $\eta (\xi )$ lead to the last expression are given by
\begin{eqnarray} 
\Lambda ^{\dag }(\xi )=-\frac{\partial }{\partial \xi }+\mathcal{W(\xi )}%
\text{ \ \ and \ \ }\Lambda (\xi )=\frac{\partial }{\partial \xi }+\mathcal{%
W(\xi )},
\end{eqnarray} %
where $\mathcal{W(\xi )}$ is the associated superpotential. In other words, when the operator $%
\Lambda ^{\dag }\Lambda $ acts on $\eta $, it leads to the following expression
\begin{eqnarray}
\Lambda ^{\dag }\Lambda \eta (\xi )=-\eta (\xi )_{\xi \xi }+\left[ \mathcal{%
W(\xi )}^{2}-\mathcal{W}(\xi )_{\xi }\right] \eta (\xi )\,.  \label{ecW}
\end{eqnarray}%
Comparing (\ref{ecJ}) with (\ref{ecW}) and remembering that $\xi
^{\prime} =g^{-1/2}$, we find the close relationship between the
superpotentials $\mathcal{W}$ and $\mathcal{J}$: 
\begin{eqnarray} 
\mathcal{W(\xi )=J(}w\mathcal{)}.
\end{eqnarray} 

Thus if $\mathcal{J(}w\mathcal{)}$ exists then $\mathcal{W(\xi )}$ exists likewise, and
the eigenvalue $\lambda $ is positive definite. In fact, we should prove that $%
\mathcal{J(}w\mathcal{)}$ exists, by finding $\mathcal{J}$ satisfying the equation
\begin{eqnarray}
\mathcal{J}^{2}-\sqrt{1-\mathcal{A}}\mathcal{J}^{\prime} =c(w)-\frac{1}{%
2\left( \mathcal{A-}1\right) }b(w),  \label{J1}
\end{eqnarray}%
with%
\begin{eqnarray}
c(w) &=&\frac{1}{4}\mathcal{A}^{\prime\prime} +\frac{7}{2}%
f^{\prime\prime} -4f^{{\prime }2}-\frac{1}{2}\mathcal{E}%
_{1}^{\prime} +2\mathcal{E}_{2} -4K, \\
b(w) &=&\frac{1}{2}f^{{\prime }2}+\frac{1}{2}\mathcal{E}_{1}^{2}+\frac{3}{8}%
\mathcal{A}^{{\prime }2}+f^{\prime} \mathcal{E}_{1}-\mathcal{A}%
^{\prime} \left( f^{\prime} +\mathcal{E}_{1}\right) .
\end{eqnarray}%
Inspired by the form of the operator $\mathcal{J}$ given by Eq. (\ref{JZ}) in the canonical case, we can define it in terms of the auxiliary field $\mathcal{Z}$ as%
\begin{eqnarray} 
\mathcal{J}=-g^{1/2}\frac{\mathcal{Z}^{\prime} }{\mathcal{Z}},
\end{eqnarray} %
for the non-canonical expression%
\begin{eqnarray} 
-gF^{\prime\prime} +\mathcal{V}F=\lambda F,
\end{eqnarray} %
where we now have the relation (\ref{J1}) for the operator $\mathcal{J}$. Thus, this expression in terms of $\mathcal{Z}$ reads
\begin{eqnarray} 
\mathcal{J}^{2}-g^{1/2}\mathcal{J}^{\prime} =g\frac{\mathcal{Z}^{^{\prime
\prime }}}{\mathcal{Z}}+\frac{1}{2}g^{\prime} \frac{\mathcal{Z}^{{\prime
}}}{\mathcal{Z}},
\end{eqnarray} %
such that when $\mathcal{A}\rightarrow 0$ ($g\rightarrow 1$) we recover (\ref%
{JZ}) and (\ref{VZ}), i.e.%
\begin{eqnarray}
\mathcal{J} &=&-g^{1/2}\frac{\mathcal{Z}^{\prime} }{\mathcal{Z}}%
\quad \longrightarrow \quad -\frac{\mathcal{Z}^{\prime} }{\mathcal{Z}} \\
\mathcal{J}^{2}-g^{1/2}\mathcal{J}^{\prime}  &=&g\frac{\mathcal{Z}%
^{\prime\prime} }{\mathcal{Z}}+\frac{1}{2}g^{\prime} \frac{\mathcal{Z}%
^{\prime} }{\mathcal{Z}} \quad \longrightarrow \quad \mathcal{J}^{2}-\mathcal{J}^{{\prime
}}=\frac{\mathcal{Z}^{\prime\prime} }{\mathcal{Z}}.
\end{eqnarray}%
Now, in order to transform this equation into the canonical form, we must suppose that%
\begin{eqnarray} 
\mathcal{Z=}\left( \mathcal{A}-1\right) \mathcal{X},
\end{eqnarray} %
which leads us to the result%
\begin{eqnarray} 
\mathcal{J}^{2}-g^{1/2}\mathcal{J}^{\prime} =-\left[ \mathcal{A}%
^{\prime\prime} \mathcal{+}2\mathcal{\mathcal{A}^{\prime} }\frac{%
\mathcal{\mathcal{X}^{\prime} }}{\mathcal{X}}\mathcal{+}\left( \mathcal{A}%
\!-\!1\right) \frac{\mathcal{X}^{\prime\prime} }{\mathcal{X}}\right] -\frac{1%
}{2\left( \mathcal{A}-1\right) }\left\{ \mathcal{A}^{\prime} \left[ 
\mathcal{A}^{\prime} \mathcal{+}\left( \mathcal{A}-1\right) \frac{%
\mathcal{\mathcal{X}^{\prime} }}{\mathcal{X}}\right] \right\}.
\end{eqnarray} %
Then comparing to (\ref{J1}), we obtain%
\begin{eqnarray}
\mathcal{A}^{\prime\prime} \mathcal{+}2\mathcal{\mathcal{A}^{\prime} }%
\frac{\mathcal{\mathcal{X}^{\prime} }}{\mathcal{X}}\mathcal{+}\left( 
\mathcal{A}-1\right) \frac{\mathcal{X}^{\prime\prime} }{\mathcal{X}}
&=&-c, \\
\mathcal{A}^{{\prime }2}\mathcal{+}\left( \mathcal{A}-1\right) \mathcal{A}%
^{\prime} \frac{\mathcal{\mathcal{X}^{\prime} }}{\mathcal{X}} &=&b,
\end{eqnarray}%
such that combining the above expressions we get%
\begin{eqnarray}
\mathcal{X}^{\prime\prime} +\mathcal{Y(}w\mathcal{)X}=0,  \label{J3}
\end{eqnarray}%
with $\mathcal{Y}$ given by%
\begin{eqnarray} 
\mathcal{Y=}\frac{1}{\mathcal{A}-1}\left[ \mathcal{A}^{\prime\prime} %
\mathcal{+}\frac{2}{\mathcal{A}-1}\left( b-\mathcal{A}^{{\prime }2}\right)
+c\right] .
\end{eqnarray} 

Thus the final form for the superpotential is%
\begin{eqnarray}
\mathcal{J}=\sqrt{1-\mathcal{A}}\left[ \frac{\mathcal{A}^{\prime} }{1-%
\mathcal{A}}-\frac{\mathcal{X}^{\prime} }{\mathcal{X}}\right] ,
\label{RP}
\end{eqnarray}
that must exist, since (\ref{J3}) has a solution. This form of the superpotential leads the Schr\"{o}dinger-like equation to its supersymmetric form (\ref{PiPit}) and implies that the corresponding spectrum of scalar fluctuations is positive definite.

Thus, this derivation actually proves that the considered tachyonic braneworld field configurations are stable under the whole sector of scalar perturbations and therefore are viable from the physical point of view.

\section{Summary and discussion} 

Braneworld models are supposed to be stable under small perturbations of all the fields that give rise to the higher dimensional field configuration. Since the braneworld we have considered is generated by gravity coupled to a bulk tachyonic scalar field, in order to be physically consistent, it is extremely important the whole model to remain stable under the action of small fluctuations of fields of both scalar and tensor nature. It is well known in the literature that the vector gravitational modes decouple from the tensor and scalar sectors and hence its influence on the brane configuration can be considered as a separate effect. Moreover, once a braneworld model proves to be stable under tensor and scalar fluctuations, it usually remains stable under the vector sector of perturbations as well \cite{Giovannini}.

In this paper we have accomplished the rather nontrivial task of proving that the tachyonic thick braneworlds are stable under the whole sector of scalar fluctuations following the method presented in \cite{Giovannini} for braneworld systems. This approach makes use of auxiliary supersymmetric potentials to define a positive definite spectrum of perturbations and shows, therefore, that there are no unstable modes with $m^2<0$.

Thus, the aforementioned expanding braneworld has already passed several consistency tests: it has proved to localise 4D gravity and to yield the Newton's law in the thin brane limit \cite{German}. In fact, in Ref. \cite{German}  some of us studied a perturbed tachyonic scalar field coupled to scalar modes of the metric fluctuations in the linear approximation. Thereat the slow-roll approximation had been used, where it lacked a rigorous and formal analysis here accomplished. That approach investigated the stability of the brane in the limit of small gradient for the tachyonic scalar field, by regarding the dynamics of scalar perturbations when the brane back reaction was taken into account. Our results here in particular provide a thorough framework whose limiting case is that one of \cite{German}. The corrections to Newton's law arising from extra-dimensional effects are exponentially suppressed and in accord with previous results reported in the literature.

 Moreover, it also localises the Standard Model fields on the 3-brane, massive scalars, gauge boson fields and fermions \cite{scalvect,fermions}, and recovers the Coulomb's law in the non relativistic Yukawa interaction of bulk gauge bosons and fermions localised on the brane. Indeed, the corrections that these laws receive from the higher dimensional realm are exponentially suppressed due to the presence of the mass gap in the corresponding spectra of the KK excitations modes. 

Finally, we would like to point that it remains to study the physical implications that our  model can present for cosmology, at least concerning the eras when the expansion of the Universe is accelerated \cite{odintsov,odintsov2}.
In fact,  tachyonic
inflation can be acquired in the context of a 5D 
AdS braneworld scenario 
 \cite{bento} and the WMAP  results for a braneworld tachyonic model of inflation implies strict bounds on the parameters in the model \cite{bento1}. Moreover, the spectrum of  braneworld inflation engendered
by a cosmological tachyon has been studied in the slow-roll approximation, where braneworld and
tachyon non-Gaussianities were shown to be  subdominant, regarding the post-inflationary
contribution \cite{calca}. 
Furthermore, the employment of tachyonic matter can  explain inflation and supplies cosmological
dark matter at late cosmological times \cite{bertolami,guo,liddle,sami}. In this context the curvaton reheating mechanisms in
tachyonic inflationary Universe models can be further  explored in the framework of braneworld  cosmology, by following similar developments as the one presented in  \cite{liddle,campu}. Usually in the first step of the expansion of the Universe the dynamics of the tachyon field in
braneworld cosmology is set out if the slow-roll over procedure is taken into account \cite{sami}.  As a closing remark, exact solutions of slow roll equations in usual 4D Friedmann-Robertson-Walker cosmology were found in \cite{sami}, presenting questions regarding the reheating in the context of the tachyonic model. We expect that effects of tachyonic potentials in the context of brane world cosmology can be further obtained and discussed using our procedure here developed, in particular in the context of more general types of thick braneworlds \cite{alex,menezes,Dutra:2013jea}.

\appendix
\section{Basic fluctuations}
Let us consider the following scalar sector of the 5D perturbed metric %
\begin{equation}
ds_{5}^{2}=e^{2f}\left[ \left( 1+2\phi \right) dw^{2}+\left( 1+2\psi \right)
\gamma _{\mu \nu }dx^{\mu }dx^{\nu }\right] .
\end{equation}%
The fluctuations can be represented in matrix form as follows%
\begin{equation}
\left( \delta g_{AB}\right) =2e^{2f}%
\begin{pmatrix}
\psi \left( \gamma _{\mu \nu }\right)  & \mathbb{O}_{4\times 1} \\ 
\mathbb{O}_{1\times 4} & \phi 
\end{pmatrix}%
,
\end{equation}%
where $\delta g^{AB}=-\mathring{g}^{AC}\mathring{g}^{BD}\delta g_{CD}$, where the symbol $\mathring{}$ indicates background components of the metric and the connection. The the explicit form for the components of the perturbed connection reads
\begin{eqnarray}
\delta \Gamma ^{w}{}_{ww} &=&\phi ^{{\prime }},\text{ \ \ \ \ \ \ \ \ \ \ \
\ }\delta \Gamma ^{\mu }{}_{\nu w}=\psi ^{{\prime }}\delta _{\nu }^{\mu },%
\text{\ \ \ \  \ \ \ \ }\delta \Gamma ^{w}{}_{w\mu }=\partial _{\mu }\phi ,
\\ \notag
\text{ }\delta \Gamma ^{w}{}_{\mu \nu } &=&\left[ 2f^{{\prime }}\left( \phi
-\psi \right) -\psi ^{{\prime }}\right] \gamma _{\mu \nu },\text{ \ \ \ \ \
}\delta \Gamma ^{\mu }{}_{ww}=-\gamma ^{\mu \nu }\partial _{\nu }\phi ,
\\ \notag
\qquad\text{  }\delta \Gamma ^{\mu }{}_{\nu \rho }&=&\delta _{\rho }^{\mu
}\partial _{\nu }\psi +\delta _{\nu }^{\mu }\partial _{\rho }\psi -\gamma
_{\nu \rho }\gamma ^{\mu \sigma }\partial _{\sigma }\psi ,
\end{eqnarray}%
while the background components of the connection are given by%
\begin{eqnarray}
\mathring{\Gamma}^{w}{}_{ww} = f^{{\prime }},\text{ \ \ \ \ \ }%
\mathring{\Gamma}^{w}{}_{\mu \nu }=-f^{{\prime }}\gamma _{\mu \nu },%
\text{ } \qquad
\mathring{\Gamma}^{w}{}_{w\mu } =0=\mathring%
{\Gamma }^{\mu }{}_{ww}, 
\qquad 
\mathring{\Gamma}^{\mu }{}_{\nu w} = f^{{\prime }}\delta _{\nu}^{\mu }.
\end{eqnarray}%

\section{Perturbed equation for the tachyonic scalar field}

In this section we shall derive in detail the expression (\ref{perttT}) for the perturbed tachyonic scalar field equation.
Let us start with the fluctuated equation for the tachyonic field 
\begin{equation}
\delta \square T-\delta \left[ \frac{\nabla _{C}\nabla _{D}T\nabla
^{C}T\nabla ^{D}T}{1+\nabla _{A}T\nabla ^{A}T}\right] =\delta \left[ \frac{1%
}{V}\partial _{\,T}V\right] .  \label{fet}
\end{equation}%
Here we have respectively the following relations%
\begin{eqnarray}
e^{2f}\delta \square T&=& \delta T^{{\prime \prime }}\!+\!3f^{{\prime
}}\delta T^{{\prime }}\!-\!2\phi T_{0}^{{\prime \prime }}\!+\!\left( 4\psi
^{{\prime }}\!-\!\phi ^{{\prime }}\!-\!6f^{{\prime }}\phi \right) T_{0}^{{\prime
}}\!+\!\overset{\gamma }{\nabla }^{\mu }\overset{\gamma }{%
\nabla }_{\mu }\delta T,   \label{fbox}\\
\delta \left[ \frac{\nabla _{C}\nabla _{D}T\nabla ^{C}T\nabla ^{D}T}{%
1+\nabla _{A}T\nabla ^{A}T}\right]&=&\frac{\delta \left[ \nabla _{C}\nabla
_{D}T\nabla ^{C}T\nabla ^{D}T\right] }{1+\nabla _{A}T_{0}\nabla ^{A}T_{0}}%
\! \notag \\ &+& \!\nabla _{C}\nabla _{D}T_{0}\nabla ^{C}T_{0}\nabla ^{D}T_{0}\delta \left[ 
\frac{1}{1+\nabla _{A}T\nabla ^{A}T}\right],   \label{fcta}\\
\delta \left( V^{-1}\partial\,_{T}V\right) &=&V^{-1}\left[ \partial\,
_{T_{0}^{2}}^{2}V-V^{-1}\left( \partial\,_{T_{0}}V\right) ^{2}\right]
\delta T .  \label{fp}
\end{eqnarray}
In (\ref{fcta}) we have the unperturbed terms%
\begin{eqnarray}
\nabla _{C}\nabla _{D}T_{0}\nabla ^{C}T_{0}\nabla ^{D}T_{0} &=&e^{-4f}\left(
T_{0}^{{\prime \prime }}-f^{{\prime }}T_{0}^{{\prime }}\right)
T_{0}^{{\prime }2}, \\
\nabla _{A}T_{0}\nabla ^{A}T_{0} &=&e^{-2f}T_{0}^{{\prime }2},
\end{eqnarray}%
while for the perturbed terms we must first consider%
\begin{eqnarray}
\delta\!\left[\nabla_{A}\nabla_{B}T\nabla^{A}T\nabla^{B}T\right] 
\!&=&\!\delta\left(g^{AC}g^{BD}\right) \nabla_{A}\nabla_{B}T_{0}
\nabla_{C}T_{0}\nabla_{D}T_{0}\! \notag \\ && +\!\mathring{g}^{AC}\mathring{g}%
^{BD}\delta \left(\nabla_{A}\nabla_{B}T\nabla_{C}T\nabla_{D}T\right) ,
\label{A5}
\\
\delta\!\left[\frac{1}{1+\nabla _{A}T\nabla ^{A}T}\right]  &=&-\frac{\delta
g^{AB}\partial _{A}T_{0}\partial _{B}T_{0}+\mathring{g}^{AB}\delta
\left( \partial _{A}T\partial _{B}T\right) }{\left( 1+\nabla _{B}T_{0}\nabla
^{B}T_{0}\right) ^{2}}.
\label{A6}
\end{eqnarray}%
We can rewrite (\ref{A5}) as%
\begin{equation}
\!\!\!\!\!\delta \left[ \nabla _{A}\nabla _{B}T\nabla ^{A}T\nabla ^{B}T\right]
\!=\!e^{-4f}T_{0}^{{\prime }}\left\{ \left[ \delta T^{{\prime \prime
}}\!-\!4\phi \left( T_{0}^{{\prime
\prime }}\!-\!f^{{\prime }}T_{0}^{{\prime }}\right) \!-\!3f^{{\prime }}\delta T^{{\prime }}\!-\!\phi ^{{\prime }}T_{0}^{{\prime }}%
\right] T_{0}^{{\prime }}+2T_{0}^{{\prime \prime }}\delta T^{{\prime
}}\right\} .
\end{equation}%
On the other hand, by calculating the equation (\ref{A6}) we arrive at the following result%
\begin{equation}
\delta \left[ \frac{1}{1+\nabla _{A}T\nabla ^{A}T}\right] =-\frac{%
2e^{-2f}T_{0}^{{\prime }}\left( \delta T^{{\prime }}-\phi T_{0}^{{\prime
}}\right) }{1+2e^{-2f}T_{0}^{{\prime }2}+e^{-4f}T_{0}^{{\prime }4}}.
\end{equation}%
By substituting the results obtained previously in (\ref{fcta}) we have%
\begin{eqnarray}
\delta \left[ \frac{\nabla _{C}\nabla _{D}T\nabla ^{C}T\nabla ^{D}T}{%
1+\nabla _{A}T\nabla ^{A}T}\right]  &=&\frac{e^{-4f}T_{0}^{{\prime
}}\left\{ \left[ -4\phi \left( T_{0}^{{\prime \prime }}-f^{{\prime
}}T_{0}^{{\prime }}\right) +\delta T^{{\prime \prime }}-3f^{{\prime
}}\delta T^{{\prime }}-\phi ^{{\prime }}T_{0}^{{\prime }}\right]
T_{0}^{{\prime }}+2T_{0}^{{\prime \prime }}\delta T^{{\prime }}\right\} }{%
1+e^{-2f}T_{0}^{{\prime }2}}  \notag \\
&&-\frac{2e^{-6f}T_{0}^{{\prime }3}\left( \delta T^{{\prime }}-\phi
T_{0}^{{\prime }}\right) \left( T_{0}^{{\prime \prime }}-f^{{\prime
}}T_{0}^{{\prime }}\right) }{1+2e^{-2f}T_{0}^{^{\prime
}2}+e^{-4f}T_{0}^{{\prime }4}}.  \label{fct}
\end{eqnarray}%
Combining (\ref{fet}), (\ref{fbox}), (\ref{fp}) and (\ref{fct}), we obtain
an expression for the perturbed tachyonic field equation
\begin{eqnarray}
&&\delta T^{{\prime \prime }}+3f^{{\prime }}\delta T^{{\prime }}-2\phi
T_{0}^{{\prime \prime }}+\left( 4\psi ^{{\prime }}-\phi ^{{\prime
}}-6f^{{\prime }}\phi \right) T_{0}^{{\prime }}+\gamma ^{\mu \nu }\overset{%
\gamma }{\nabla }_{\mu }\overset{\gamma }{\nabla }_{\nu }\delta T  \notag \\
&&+\frac{e^{-2f}T_{0}^{{\prime }}\left\{ \left[ 4\phi \left(
T_{0}^{{\prime \prime }}-f^{{\prime }}T_{0}^{{\prime }}\right) -\delta
T^{{\prime \prime }}+3f^{{\prime }}\delta T^{{\prime }}+\phi ^{{\prime
}}T_{0}^{{\prime }}\right] T_{0}^{{\prime }}-2T_{0}^{^{\prime \prime
}}\delta T^{{\prime }}\right\} }{1+e^{-2f}T_{0}^{{\prime }2}}  \label{fteA} \\
&&+\frac{2e^{-4f}T_{0}^{{\prime }3}\left( \delta T^{{\prime }}-\phi
T_{0}^{{\prime }}\right) \left( T_{0}^{{\prime \prime }}-f^{{\prime
}}T_{0}^{{\prime }}\right) }{1+2e^{-2f}T_{0}^{^{\prime
}2}+e^{-4f}T_{0}^{{\prime }4}}  \notag 
=e^{2f}V^{-1}\left[ \partial\,_{T_{0}^{2}}^{2}V-V^{-1}\left( \partial\,
_{T_{0}}V\right) ^{2}\right] \delta T.  
\end{eqnarray}
Finally, by applying the gauge relations%
\begin{equation}
\phi =-2\psi \text{ \ \ and \ \ }\delta T=\mathcal{F}\left( \psi ^{{\prime
}}+2f^{{\prime }}\psi \right) ,
\end{equation}%
and with the aid of Eq. (\ref{mata}) we can rewrite Eq. (\ref{fteA}) as follows%
\begin{eqnarray}
&&\left\{ \mathcal{A-}1\right\} \delta T^{{\prime \prime }}+\left\{
f^{{\prime }}\left( 2\mathcal{A}^{2}\!-\!3\mathcal{A-}3\right)\!+\!2\frac{%
T_{0}^{{\prime \prime }}}{T_{0}^{{\prime }}}\mathcal{A}\left(1\!-\!\mathcal{A}%
\right)\right\}\!\delta T^{{\prime }}\! \notag \\ && +\!\left\{ \frac{e^{2f}}{V}\left[ \partial\,_{T^{2}}^{2}V\!-\!V^{-1}\left(
\partial\,_{T}V\right) ^{2}\right]\!-\!\frac{4}{\mathcal{F}}T_{0}^{{\prime
}}\right\}\!\delta T  \notag
\\
&&-4\left\{ \left( \mathcal{A}^{2}-2\mathcal{A}+1\right) T_{0}^{^{\prime
\prime }}-f^{{\prime }}\left( \mathcal{A}^{2}-2\mathcal{A}-1\right)
T_{0}^{{\prime }}\right\} \psi   
=\gamma ^{\mu \nu }\overset{\gamma }{\nabla }_{\mu }\overset{\gamma }{%
\nabla }_{\nu }\delta T.
\end{eqnarray}

\section{Derivation of the Schr\"{o}dinger-like equation}
Let us now consider the expression (\ref{eqpsi}) for the perturbed energy-momentum tensor of the tachyonic scalar field%
\begin{equation}
\delta T^{\mathcal{B}}{}_{w}^{w}=\delta{T}^{\mathcal{B}%
}{}_{\nu }^{\mu }+\frac{e^{-2f}T_{0}^{{\prime }}}{\sqrt{1+e^{-2f}T_{0}^{^{%
\prime }2}}}\left[ T_{0}^{{\prime }}\left( \partial _{T_{0}}V\right) \delta
T-\frac{e^{-2f}T_{0}^{{\prime }2}V\left( \delta T^{{\prime }}+2\psi
T_{0}^{{\prime }}\right) }{1+e^{-2f}T_{0}^{{\prime }2}}+2V\left( \delta
T^{{\prime }}+2\psi T_{0}^{{\prime }}\right) \right] ,
\end{equation}%
where $\mu=\nu$ in the first term of the right hand side with no summation regarding these indices.
Combining it with Eqs. (\ref{e1}) and (\ref{e3}) and remembering that $\phi =-2\psi $, we get%
\begin{equation}
\psi ^{{\prime \prime }}+f^{{\prime }}\psi ^{{\prime }}+\left(
4f^{{{\prime \prime }}}-4f^{{\prime }2}\right) \psi -\mathcal{T}=\left(
\gamma ^{\mu \nu }\overset{\gamma }{\nabla }_{\mu }\overset{\gamma }{\nabla }%
_{\nu }+2K\right) \psi ,  \label{aux1}
\end{equation}%
with $\mathcal{T}$ being given by%
\begin{equation}
\mathcal{T=}\frac{1}{\mathcal{F}}\left[ T_{0}^{{\prime }}\left(
V^{-1}\partial _{T_{0}}V\right) \delta T+2\left( \delta T^{{\prime }}+2\psi
T_{0}^{{\prime }}\right) -\frac{e^{-2f}T_{0}^{{\prime }2}\left( \delta
T^{{\prime }}+2\psi T_{0}^{{\prime }}\right) }{1+e^{-2f}T_{0}^{{\prime }2}%
}\right] .
\end{equation}%
We can rewrite the first two terms of this sum as follows%
\begin{eqnarray}
T_{0}^{{\prime }}\left( V^{-1}\partial _{T_{0}}V\right) \delta T &=&%
\mathcal{F}T_{0}^{{\prime }}\left( \psi ^{{\prime }}+2f^{{\prime }}\psi
\right) V^{-1}\partial _{T_{0}}V, \\
\delta T^{{\prime }}+2\psi T_{0}^{{\prime }} &=&\mathcal{F}\left[ \psi
^{{\prime \prime }}+\left( \frac{\mathcal{F}^{{\prime }}}{\mathcal{F}}%
+2f^{{\prime }}\right) \psi ^{{\prime }}+2\left( \frac{\mathcal{F}%
^{{\prime }}}{\mathcal{F}}f^{{\prime }}+f^{{{\prime \prime }}}+\frac{%
T_{0}^{{\prime }}}{\mathcal{F}}\right) \psi \right] ,
\end{eqnarray}%
where we have used the relation $\delta T=\mathcal{F}\left( \psi ^{{\prime
}}+2f^{{\prime }}\psi \right) $. Hence it yields 
\begin{eqnarray}
\mathcal{T} &=& \left( 2-\frac{e^{-2f}T_{0}^{{\prime }2}}{
1+e^{-2f}T_{0}^{{\prime }2}}\right) \psi ^{{\prime \prime }}+\left[
T_{0}^{{\prime }}V^{-1}\partial _{T_{0}}V+\left( \frac{\mathcal{F}
^{{\prime }}}{\mathcal{F}} \notag +2f^{{\prime }}\right) \left( 2-\frac{
e^{-2f}T_{0}^{{\prime }2}}{1+e^{-2f}T_{0}^{{\prime }2}}\right) \right]
\psi ^{{\prime }} \\
&&+2\left[ f^{{\prime }}T_{0}^{{\prime }}V^{-1}\partial _{T_{0}}V+\left( 
\frac{\mathcal{F}^{{\prime }}}{\mathcal{F}}f^{{\prime }}+f^{{{\prime
\prime }}}+\frac{T_{0}^{{\prime }}}{\mathcal{F}}\right) \left( 2-\frac{
e^{-2f}T_{0}^{{\prime }2}}{1+e^{-2f}T_{0}^{{\prime }2}}\right) \right]
\psi .
\end{eqnarray}
Putting this last expression into (\ref{aux1}) leads to the following relation%
\begin{eqnarray}
&&\left[ \mathcal{A}-1\right] \psi ^{{\prime \prime }}+\left[ f^{{\prime
}}-\mathcal{D}+\mathcal{B}\left( \mathcal{A-}2\right) \right] \psi
^{{\prime }}+2\left[ 2f^{{{\prime \prime }}}-2f^{{\prime }2}-f^{{\prime
}}\mathcal{D}+\mathcal{C}\left( \mathcal{A-}2\right) - 4K \right] \psi   \notag \\
&=&\left( \gamma ^{\mu \nu }\overset{\gamma }{\nabla }_{\mu }\overset{\gamma 
}{\nabla }_{\nu }-2K\right) \psi ,  \label{aux2}
\end{eqnarray}%
which is equivalent to Eq. (\ref{e4}); here the functions $\mathcal{A}$, $\mathcal{B}$, $\mathcal{C}$ and $%
\mathcal{D}$ have been introduced in Eq. (\ref{mata}) and define the following quantities
\begin{equation}
\mathcal{E}_{1}\mathcal{=}-\mathcal{D}+\mathcal{B}\left( \mathcal{A-}%
2\right) \text{ \ \ and \ \ }\mathcal{E}_{2}=-f^{{\prime }}\mathcal{D}+%
\mathcal{C}\left( \mathcal{A-}2\right) - 4K ,
\end{equation}%
such that Eq. (\ref{aux2}) becomes%
\begin{equation}
\left[ \mathcal{A}-1\right] \psi ^{{\prime \prime }}+\left[ f^{{\prime }}+%
\mathcal{E}_{1}\right] \psi ^{{\prime }}+2\left[ 2f^{{{\prime \prime
}}}-2f^{{\prime }2}+\mathcal{E}_{2}\right] \psi =\left( \gamma ^{\mu \nu }%
\overset{\gamma }{\nabla }_{\mu }\overset{\gamma }{\nabla }_{\nu }-2K\right)
\psi .  \label{aux2b}
\end{equation}%

In order to transform the equation (\ref{aux2b}) into a Schr\"{o}dinger-like equation,
we introduce the auxiliary function $\alpha$ as follows 
\begin{equation}
\left( \mathcal{A-}1\right) \frac{\alpha ^{{\prime }}}{\alpha }=f^{{\prime
}}+\mathcal{E}_{1}\text{ \ \ and \ \ }\psi =\frac{1}{\alpha ^{1/2}(w)}%
F\left( w,x^{\mu }\right) .  \label{aux3}
\end{equation}%
Therefore, it 
leads us to the following expression for Eq. (\ref{aux2b})%
\begin{eqnarray}
&& \!\!\!\!\!\left[ \mathcal{A-}1\right] F^{{\prime \prime }}+\left[ 4\left(f^{{{\prime
\prime }}}-f^{{\prime }2}\right)+2\mathcal{E}_{2}+\frac{1}{2}\left( \mathcal{A-}%
1\right) \left( \frac{1}{2}\frac{\alpha ^{{\prime }2}}{\alpha ^{2}}-\frac{%
\alpha ^{{{\prime \prime }}}}{\alpha }\right) -4K \right] F \notag \\ &=& \left( \overset{\gamma }{\nabla }^{\mu }\overset{\gamma 
}{\nabla }_{\mu }-2K\right) F. \label{aux4}
\end{eqnarray}%
With the above introduced definition for $\alpha$ (\ref{aux3}) and taking into account the expression for 
$(\alpha ^{-1}\alpha ^{{\prime}})^{{\prime }}$, we have the following equality%
\begin{equation}
\frac{\alpha ^{{\prime \prime }}}{\alpha }=\frac{1}{\mathcal{A-}1}\left[
f^{{\prime \prime }}+\mathcal{E}_{1}^{{\prime }}-\frac{\mathcal{A}%
^{{\prime }}\left( f^{{\prime }}+\mathcal{E}_{1}\right) }{\mathcal{A-}1}%
\right] +\frac{\alpha ^{{\prime }2}}{\alpha ^{2}}.
\end{equation}%
Therewith, by replacing this expression in (\ref{aux4}), we arrive at a Schr\"{o}dinger-like equation%
\begin{eqnarray}
-\left(1 - \mathcal{A}\right) F^{\prime\prime} +\mathcal{V}F=
m^2 F,  \label{etsA}
\end{eqnarray}
with the corresponding potential
\begin{eqnarray}
\label{potlA}
\mathcal{V} &\mathcal{=}&4f^{{{\prime \prime }}}-4f^{{\prime }2}+2%
\mathcal{E}_{2}+\frac{1}{2}\left( \mathcal{A-}1\right) \left( \frac{1}{2}%
\frac{\alpha ^{{\prime }2}}{\alpha ^{2}}-\frac{\alpha ^{{{\prime \prime
}}}}{\alpha }\right)  \notag \\
&=&\frac{7}{2}f^{{{\prime \prime }}}-4f^{{\prime }2}-\frac{1}{2}\mathcal{E%
}_{1}^{{\prime }}+2\mathcal{E}_{2}-\frac{1}{2\left( \mathcal{A-}1\right) }%
\left[ \frac{1}{2}f^{{\prime }2}+\frac{1}{2}\mathcal{E}_{1}^{2}+f^{{\prime
}}\mathcal{E}_{1}-\mathcal{A}^{{\prime }}\left( f^{{\prime }}+\mathcal{E}%
_{1}\right) - 4K \right] , \notag \\  \label{potA}
\end{eqnarray}%
where (\ref{etsA}) and (\ref{potA}) are our looked for expressions (\ref{ets}) and (\ref{potl}), respectively.

\acknowledgments

AMK is grateful to CAPES and Programa Ci\^{e}ncia sem Fronteiras (CsF) for financial support and to the ICF, UNAM, for hospitality. AHA thanks the ICF, UNAM, UAM-Iztapalapa and ICTP-SAIFR for hospitality and is grateful as well to VIEP-BUAP-HEAA-EXC15-I and PRODEP NPT-411. GG, AHA and DMM acknowledge SNI and support from the grant PAPIIT-UNAM, IN103413-3, ``Teor\'{i}as de Kaluza-Klein, inflaci\'{o}n y perturbaciones gravitacionales''. RdR thanks CNPq Grants No. 303027/2012-6 and No. 473326/2013-2, and FAPESP No. 2015/10270-0, for partial financial support.

\end{document}